\documentclass[letterpaper,journal]{IEEEtran}  
\usepackage{amsmath,amsfonts}
\usepackage{algorithmic}
\usepackage{algorithm}
\usepackage{array}
\usepackage[caption=false, font=normalsize, labelfont=rm]{subfig} 
\usepackage{textcomp}
\usepackage{stfloats}
\usepackage{url}
\usepackage{verbatim}
\usepackage{graphicx}
\usepackage{cite}
\usepackage{amssymb}

\usepackage{amsthm}
\newtheorem{theorem}{Theorem}
\usepackage{xcolor}
\usepackage{epstopdf}
\usepackage{hyperref}            

\begin{document}

\title{BEM-Assisted Low-Complexity Channel Estimation for AFDM Systems over Doubly Selective Channels}

\author{
 Limin Liu, 
 Zhe Li, 
 Qihao Peng,~\IEEEmembership{Member,~IEEE,} 
 Qu Luo, ~\IEEEmembership{Member,~IEEE,} 
 Pei Xiao, ~\IEEEmembership{Senior Member,~IEEE,}  
 Haowei Wu,  ~\IEEEmembership{Member,~IEEE} 
       
\thanks{ Limin Liu is with the School of Microelectronics and Communication Engineering, Chongqing University, Chongqing 400044, China. (email: liulimin4858@163.com). Zhe Li is with the Research and Development Department, China Academy of Launch Vehicle Technology, Beijing, China. (email: zheli@163.com). Qihao Peng, Qu  Luo,  Pei Xiao and are  with the  5G \& 6G  Innovation Centre, University of Surrey, U. K. (email: \{q.peng, q.u.luo,  p.xiao\}@surrey.ac.uk). Haowei Wu is with the School of Microelectronics and Communication Engineering, Chongqing University, Chongqing 400044, China, and also with the Chongqing Key Laboratory of Space Information Network and Intelligent Information Fusion. (email: wuhaowei@cqu.edu.cn). Corresponding author (Haowei Wu and Qu Luo).}
}

\markboth{  }%
{ }

 \maketitle

\begin{abstract}
In this paper, we propose a low-complexity channel estimation scheme of affine frequency division multiplexing (AFDM)  based on generalized complex exponential basis expansion model (GCE-BEM) over doubly selective channels. The  GCE-BEM is used to solve fractional Doppler dispersion.
Then, the closed-form expression of channel estimation error is derived for the minimum mean square error (MMSE) estimation algorithm. Based on the estimated channel, the MMSE detection is adopt to characterize the impacts of estimated channel on bit error rate (BER) by deriving the theoretical lower bound. Finally, numerical results demonstrate that the proposed scheme effectively mitigates severe inter-Doppler interference (IDoI). Our theoretical performance analysis can perfectly match the Monte-Carlo results, validating the effectiveness of our proposed channel estimation based on GCE-BEM.
\end{abstract}

\vspace{-1em}
\section{Introduction}
\IEEEPARstart{T}{he} sixth generation (6G) mobile communication technology is expected to provide ultra-reliable, low latency, and high-speed data transmission communication in high-speed mobile scenarios, mainly including unmanned aerial vehicles, vehicle-to-vehicle, high-speed rail, and low-orbit satellites \cite{ref2,ref1}. Although orthogonal frequency division multiplexing (OFDM) exhibits promising applications, it falls short to meet the high-speed mobile scene of 6G, as Doppler destroys the orthogonality between OFDM subcarriers, resulting in severe inter-carrier interference (ICI)\cite{ref3,ref4}. Therefore, it is imperative to design a new waveform system to support the future mobile communication system with high mobility.

To address this challenge, several alternative waveforms have been proposed for robust transmission over doubly selective (time- and frequency-dispersive) channels, including orthogonal chirp division multiplexing (OCDM) \cite{ref5}, and orthogonal time frequency space (OTFS) modulation \cite{ref6}.
More recently, affine frequency division multiplexing (AFDM) \cite{ref7} has garnered increasing attention for high-mobility scenarios and is applicable for integrated sensing and communications (ISAC) \cite{ref8, ref9}. It utilizes quadratic exponential waveforms i.e., chirp-modulated signals, as mutually orthogonal subcarriers \cite{ref10}. By adaptively tuning the chirp rate in accordance with the channel's Doppler profile, AFDM enables a separable and quasi-static channel representation, thereby achieving full diversity over doubly selective channels \cite{ref11}. 

Even though AFDM exhibits promising characteristics, the channel estimation is one of the fundamental challenges in AFDM systems. In \cite{ref7}, the authors proposed an embedded pilot-assisted approximate maximum likelihood (EPA-AML) channel estimation scheme. However, due to multipath propagation, EPA-AML requires an exhaustive search to estimate the channel parameters, resulting in prohibitive computational complexity. To reduce complexity, a low-complexity diagonally reconstructed channel estimation method was introduced in \cite{ref12} for multi-antenna AFDM systems. However, it relies on a large $c_1$ value to ensure that multipaths channels  are well separated in the DAFT domain, and  the accuracy of the estimation of this method depends on the selection of the threshold. In \cite{ref13}, a superimposed pilot-based channel estimation technique was proposed, which overlays multiple pilots onto data symbols to eliminate the need for a guard interval (GI). However, this approach only accounts for integer Doppler shifts and fails to consider the more general fractional Doppler effects.
Despite these advancements, designing a generalized channel estimation scheme that achieves low complexity and robust performance over doubly selective channels—particularly in the presence of strong fractional Doppler shifts—remains an open and important problem.

In this paper, to address the challenges of the fractional frequency dispersion and computational complexity of channel estimation, we propose a low-complexity embedded pilot-assisted channel estimation scheme for AFDM systems over doubly selective channels. Firstly, we approximate the AFDM signal in the discrete affine Fourier transform (DAFT) domain by using generalized complex exponential basis expansion model (GCE-BEM) \cite{ref14} and thus avoid the ambiguities caused by fractional frequency dispersion and multipath propagation. Furthermore, by adopting GCE-BEM, the number of estimated parameters can be reduced, thereby significantly reducing the complexity of channel estimation. Then, the impacts of GCE-BEM on channel estimation are characterized by deriving the closed-form expression of estimation error, revealing the relationship between the model errors and estimation errors. Based on the estimated channel, the lower bound of bit error rate (BER) is analyzed, which provides in-depth insights into the system design. Finally, simulation results validate our rigorous derivations and analysis.

The rest of this paper is organized as follows. Section \uppercase\expandafter{\romannumeral2} describes the GCE-BEM-based AFDM system model. The channel estimation and BER are analyzed in Section \uppercase\expandafter{\romannumeral3}. Numerical results are presented in Section \uppercase\expandafter{\romannumeral4}. The conclusions are drawn in Section \uppercase\expandafter{\romannumeral5}.


\section{System Model}
\subsection{AFDM System Model}

Denote $\mathbf{x}\in {{\mathbb{A}}^{N\times 1}}$ as a vector of $N$ quadrature amplitude modulation (QAM) symbols in the DAFT domain, where $\mathbb{A}$ denotes the modulation alphabet. Then, $\mathbf{x}$ is mapped to the time domain via an $N$ point IDAFT, i.e., $\mathbf s = {{\bf{A}}^H}{\bf{x}}$, where  ${\bf{A}}={{\bf{\Lambda}} _{{c_2}}}{\bf{F}}{{\bf{\Lambda }}_{{c_1}}}$ denotes the DAFT matrix, ${\bf{F}} \in {\mathbb{C}^{N \times N}}$ is the $N$-point FFT matrix, the superscript $H$ denotes complex conjugate transpose. $c_1$ and $c_2$ are the two AFDM parameters, and  
\begin{equation}
\label{deqn_ex1}
{{\bf{\Lambda}} _c} = {\rm{diag}}({e^{ - j2\pi c{n^2}}},n = 0,1, \ldots ,N - 1,c \in \{c_1, c_2\}
).
\end{equation}

We consider a doubly selectively channel with its channel response at time $n$ and delay $l$ given as $ {h_n}(l) = \sum\nolimits_{i = 1}^P {{h_i}{e^{ - j2\pi {f_i}n}}\delta (l - {l_i})}.$ 
At the receiver side, after the DAFT operation, the received vector $\mathbf{y}$ in the DAFT domain can be expressed as\cite{ref7}
\begin{equation}
\begin{aligned}
\label{deqn_ex2}
{\bf{y}}& = \sum\limits_{i = 1}^P {{h_i}{\bf{A}}{{\bf{\Gamma}} _{{\rm{CP}}{{\rm{P}}_i}}}{{\bf{\Delta}} _{{f_i}}}{{\bf{\Pi}} ^{{l_i}}}{{\bf{A}}^H}{\bf{x}}}  + {\bf{ w}} \\
&= \sum\limits_{i = 1}^P {{h_i}{{\bf{H}}_i}{\bf{x}}}  + {\bf{ w}} = {\bf{AH}}{{\bf{A}}^H}{\bf{x}} + {\bf{ w}} = {{\bf{H}}_{{\rm{eff}}}}{\bf{x}} + {\bf{w}},
\end{aligned}
\end{equation}
where ${\bf{ w}} \sim{{\cal C}{\cal N}}(0,{N_0}{\bf{I}})$ is the DAFT domain additive white Gaussian noise vector with variance $N_0$, and ${\bf{H}}$ is the time domain channel matrix \footnote{Similar to  \cite{ref7}, the chirp parameters are set to ${c_1} = \frac{{2({\alpha _{\max }} + {k_\nu }) + 1}}{{2N}}$, where ${\alpha _{\max }}$ is the   maximum integer Doppler   normalized   to the subcarrier spacing, and ${k_\nu }$ denotes a non-negative integer which captures fractional Doppler shifts. $c_2$ is chosen to be either an arbitrary irrational number or a rational number that is sufficiently less than $\frac{1}{2N}$.}.


\subsection{GCE-BEM Channel Model in AFDM System}
In this paper, the GCE-BEM is adopted to  reduce  the number of parameters to be estimated  and the overall complexity of channel estimation. By applying the GCE-BEM to the $\bf{H}$ in \eqref{deqn_ex2}, we have
\begin{equation}
\label{deqn_ex6}
{\bf{H}} = \sum\limits_{q = 0}^Q {{\rm{diag}}\{ {{\bf{b}}_q}\} {{\bf{F}}^H}{\rm{diag}}\{ {{\bf{F}}_L}{{\bf{g}}_q}\} {\bf{F}}}  + {{\bf{E}}_{\rm{mod} }},
\end{equation}
where ${{\bf{b}}_q} = [{b_q}(0),{b_q}(1), \cdots ,{b_q}(N - 1)]$ represents the $q$-th BEM basis function, ${b_q}(n) = {e^{j2\pi \frac{{(q - \left\lceil {Q/2} \right\rceil )n}}{RN}}}$ is the $q$-th BEM basis function of the $n$-th time. $R$ is the oversampling coefficient of GCE-BEM and $\left\lceil  \cdot  \right\rceil$ represents the ceiling operator. ${{\bf{g}}_q} = {[{g_q}[0],{g_q}[1], \cdots ,{g_q}[L]]^T}$ denotes the $q$-th BEM coefficient vector, and ${g_q}(l)$ denotes the $q$-th BEM coefficient of the $l$-th path. $L$ represents the maximum normalized time delay. ${{\bf{E}}_{\rm{mod} }}$ is the corresponding channel model error matrix,  and ${\bf{F}}_L$ is the first $L+1$ column of $\sqrt N {\bf{F}}$. $Q$ denotes the order of the BEM basis function which satisifies $Q \ge 2\left\lceil {R{f_{\max }}N{T_{\rm{s}}}} \right\rceil $ \cite{ref14}, ${f_{\max }}$ is the maximum Doppler frequency shift, and ${T_{\rm{s}}}$ is the sampling interval.

According to \eqref{deqn_ex6}, the number of time-varying channel coefficients is reduced from $N(L + 1)$ to $(Q + 1)(L + 1)$, thereby significantly reducing the complexity of channel estimation.
 
 By substituting \eqref{deqn_ex6} into \eqref{deqn_ex2}, the received AFDM signal $\bf{y}$ is  rewritten as
\begin{equation}
\label{deqn_ex7}
{\bf{y}} = \sum\limits_{q = 0}^Q {{\bf{A}}{\rm{diag}}\{ {{\bf{b}}_q}\} {{\bf{F}}^H}{\rm{diag}}\{ {{\bf{F}}_L}{{\bf{g}}_q}\} {\bf{F}}{{\bf{A}}^H}{\bf{x}}}  + {{\bf{z}}_{\rm{mod} }} + {\bf{ w}},
\end{equation}
where ${{\bf{z}}_{\rm{mod} }} = {\bf{A}}{{\bf{E}}_{\rm{mod} }}{{\bf{A}}^H}{\bf{x}}$ is the model error of the BEM channel model at the receiver. 


\section{CHANNEL ESTIMATION BASED ON EMBEDDED PILOT}
\subsection{Pilot Design}
\begin{figure}[htpb]
\centering
\includegraphics[width=3in]{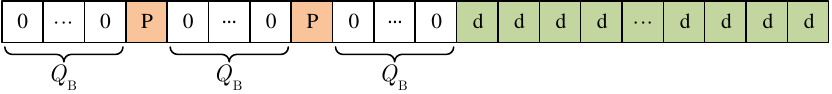}
\caption{Transmitted symbols arrangement (‘P’: pilot, ‘d’: data, ‘0’: guard).}
\label{fig_1}
\end{figure}

The embedded pilot scheme structure is illustrated in Fig. \ref{fig_1}. Specifically, two pilots, ${x_{\rm{p_1}}}$ and ${x_{\rm{p_2}}}$, are inserted into each AFDM symbol at positions $Q_{\rm{B}}$ and $2Q_{\rm{B}}+1$, respectively. Note that, $Q_{\rm{B}}$ null subcarriers are placed on both sides and between the two pilots as guard subcarriers.
According to \eqref{deqn_ex2}, the row and column indices $p$ and $q$ of the non-zero elements in the channel matrix ${{\bf{H}}_i}$ satisfy ${(p + {\rm{lo}}{{\rm{c}}_i} - {k_\nu })_N} \le q \le {(p + {\rm{lo}}{{\rm{c}}_i} + {k_\nu })_N}$, where ${\rm{lo}}{{\rm{c}}_i} \buildrel \Delta \over = {({\alpha _i} + 2N{c_1}{l_i})_N}$, ${( \cdot )_N}$ represents modular $N$ operation. This implies that the $q$-th element of the transmitted signal ${\bf{x}}$ will spread to the range $[{(q - {\rm{lo}}{{\rm{c}}_i} - {k_\nu })_N},{(q - {\rm{lo}}{{\rm{c}}_i} + {k_\nu })_N}]$ after passing through the channel. Essentially, the GCE-BEM can be interpreted as sampling the Doppler spectrum with a sampling interval of $1/RN$. Consequently, the pilot ${x_{\rm{p_1}}}$ are spread over the range $[{Q_{\rm{B}}} - 2N{c_1}{l_{\max }} - \frac{Q}{2},{Q_{\rm{B}}} + \frac{Q}{2}]$ at the receiver. Thus, we set ${Q_{\rm{B}}} = Q + 2N{c_1}{l_{\max }}$, where ${Q} \ge 4\left\lceil {{f_{\max }}N{T_{\rm{s}}}} \right\rceil$ is the BEM order when the oversampling coefficient of GCE-BEM $R=2$.

\vspace{-1em}
\subsection{Channel Estimation}
In this subsection, we delve into the detailed channel estimation algorithm design under the embedded pilot architecture. The transmitted signal in the DAFT domain, denoted as $\mathbf{x}$, can be decomposed into a pilot component and a data component, i.e., $\mathbf{x} = \mathbf{x}_{\rm{p}} + \mathbf{x}_{\rm{d}}$. 
Note that only the portion of the received signal corresponding to the pilot symbols is used for channel estimation. These symbols can be expressed as
\begin{equation}
\begin{aligned}
\label{deqn_ex11}
{{\bf{y}}_{\rm{p}}} = {{\bf{\Psi}} _{{\rm{p}},{\rm{p}}}}{\bf{g}} + {{\bf{\Psi}} _{{\rm{d}},{\rm{p}}}}{\bf{g}} + {{\bf{z}}_{{\rm{mod}} ,{\rm{p}}}} + {{\bf{w}}_{\rm{p}}},
\end{aligned}
\end{equation}
where  
${{\bf{y}}_{\rm{p}}} = {{\bf{T}}_{\rm{p}}}{\bf{y}}$, ${{\bf{\Psi}} _{{\rm{p}},{\rm{p}}}} = {{\bf{T}}_{\rm{p}}}{\bf{\Psi}}_{\rm{p}} $, ${{\bf{\Psi}} _{{\rm{d}},{\rm{p}}}} = {{\bf{T}}_{\rm{p}}}{\bf{\Psi}}_{\rm{d}} $, ${{\bf{z}}_{{\rm{mod}},{\rm{p}}}} = {{\bf{T}}_{\rm{p}}}{{\bf{z}}_{\rm{mod} }}$ and ${{\bf{w}}_{\rm{p}}} = {{\bf{T}}_{\rm{p}}}{{\bf{w}}}$ are the part of $\bf{y}$, ${\bf{\Psi}}_{\rm{p}}$, ${\bf{\Psi}}_{\rm{d}}$, ${\bf{z}}_{{\rm{mod}}}$ and ${\bf{w}}$ are related to the channel estimation, respectively. ${{\bf{T}}_{\rm{p}}} = {\left[ {{{\bf{I}}_N}} \right]_{{\rm{in}}{{\rm{d}}_{\rm{p}}}}}$, and ${{\rm{in}}{{\rm{d}}_{\rm{p}}}} = [\frac{Q}{2}:2Q_{\rm{B}} + \frac{Q}{2}+1]$ represents the index range. ${{\bf{\Psi}} _{\rm{p}}} = [{{\bf{D}}_{0,{\rm{p}}}},{{\bf{D}}_{1,{\rm{p}}}}, \ldots ,{{\bf{D}}_{Q,{\rm{p}}}}]\in {{\mathbb{C}}^{N \times (Q+1)(L + 1)}}$ is a deterministic matrix associated with the pilot,   ${{\bf{D}}_{q,{\rm{p}}}} = ({\bf{A}}{\rm{diag}}\{ {{\bf{b}}_q}\} {{\bf{F}}^{\rm{H}}}{\rm{diag}}\{{\bf{F}}{{\bf{A}}^{\rm{H}}}{{\bf{x}}_{\rm{p}}}\}){{\bf{F}}_L}$ denotes the pilot matrix. ${{\bf{\Psi}} _{\rm{d}}} = [{{\bf{D}}_{0,{\rm{d}}}},{{\bf{D}}_{1,{\rm{d}}}}, \ldots ,{{\bf{D}}_{Q,{\rm{d}}}}]\in {{\mathbb{C}}^{N \times (Q+1)(L + 1)}}$ is a random matrix associated with the data, and ${{\bf{D}}_{q,{\rm{d}}}} = ({\bf{A}}{\rm{diag}}\{ {{\bf{b}}_q}\} {{\bf{F}}^{\rm{H}}}{\rm{diag}}\{{\bf{F}}{{\bf{A}}^{\rm{H}}}{{\bf{x}}_{\rm{d}}}\}){{\bf{F}}_L}$ denotes data matrix. ${\bf{g}} = {[{{\bf{g}}_0}^T, {{\bf{g}}_1}^T, \ldots ,{{\bf{g}}_Q}^T ]^T}$ is the BEM basis coefficient vector.

\begin{theorem}
\label{theorem1}
Based on the AFDM signal in \eqref{deqn_ex11}, the minimum mean square error (MMSE) estimator of the channel coefficients can be expressed as
\begin{equation}
\begin{aligned}
\label{deqn_ex12}
{\bf{\hat g}} = {{\bf{R}}_{\rm{g}}}{\bf{\Psi}} _{{\rm{p}},{\rm{p}}}^H{({{\bf{\Psi}} _{{\rm{p}},{\rm{p}}}}{{\bf{R}}_{\rm{g}}}{\bf{\Psi}} _{{\rm{p}},{\rm{p}}}^H + {{\bf{R}}_{{\rm{d}},{\rm{p}}}} + {{\bf{R}}_{{\rm{z}},{\rm{p}}}} + {{\bf{R}}_{{\rm{w}},{\rm{p}}}})^{ - 1}}{{\bf{y}}_{\rm{p}}},
\end{aligned}
\end{equation}
where ${{\bf{R}}_{\rm{g}}} = {\bf{\Theta}}^{-1}{{\bf{R}}_{{\rm{hh}}}}({\bf{\Theta}}^H)^{-1}$ and ${{\bf{R}}_{{\rm{hh}}}}$ denote covariance matrix of BEM basis coefficient vector and channel vector, respectively. ${\bf{\Theta}} = {\bf{B}} \otimes {{\bf{I}}_L}$, and ${\bf{B}} = [{{\bf{b}}_0},{{\bf{b}}_1}, \ldots ,{{\bf{b}}_Q}]$ represents the basis function matrix. $\otimes$ denotes the Kronecker Product. ${{\bf{R}}_{{\rm{d}},{\rm{p}}}} = {{\bf{T}}_{\rm{p}}}{{\bf{R}}_{\rm{d}}}{\bf{T}}_{\rm{p}}^H$ is the data covariance matrix related to pilot, and ${{\bf{R}}_{\rm{d}}} $ is the data covariance matrix, given in \eqref{deqn_ex20}. ${{\bf{R}}_{{\rm{z}},{\rm{p}}}} = {{\bf{T}}_{\rm{p}}}{{\bf{R}}_{\rm{z}}}{\bf{T}}_{\rm{p}}^H$ is the model error covariance matrix related to the pilot. The covariance matrix ${{\bf{R}}_{\rm{z}}} $ and  BEM estimation error, i.e.,  ${{\bf{R}}_{\tilde{{\rm{g}}}}}$,   are  given by \eqref{deqn_ex22}   and \eqref{deqn_ex25}, respectively.
\end{theorem}

\emph{Proof: {\rm{Refer to Appendix A.}}} \hfill$\blacksquare$

According to \eqref{deqn_ex12}, the estimated time-domain channel matrix can be expressed as ${\bf{\hat H}} = \sum\nolimits_{q = 0}^Q {{\rm{diag}}\{ {{\bf{b}}_q}\} {{\bf{F}}^H}{\rm{diag}}\{ {{\bf{F}}_L}{{\bf{\hat g}}_q}\} {\bf{F}}}$. 
The  normalized mean square error (NMSE) of system channel estimation is given by   \eqref{deqn_ex26}.

The complexity of the proposed scheme is dominated by the operation in (\ref{deqn_ex12}), whose computational cost can be approximated as ${\cal O}({N^2}QL)$. In contrast, for traditional MMSE-based channel estimation without BEM, the complexity is  ${\cal O}({N^3})$ since the dimension of the matrix inversion is $N$.  As $  QL \ll N$, the proposed GCE-BEM AFDM system exhibits significantly lower computational complexity.

\subsection{BER Performance Analysis}
Based on the estimated channel matrix ${\bf{\hat H}}$, the received signal without pilot interference is given by 
\begin{equation}
\begin{aligned}
\label{deqn_ex15}
{\bf{\hat y}} = {\bf{y}} - {{\bf{\hat H}}_{{\rm{eff}}}}{{\bf{x}}_{\rm{p}}} = {{\bf{\hat H}}_{{\rm{eff}}}}{{\bf{x}}_{\rm{d}}} + {{\bf{\tilde H}}_{{\rm{eff}}}}{\bf{x}} + {{\bf{z}}_{\rm{mod} }} + {\bf{w}},
\end{aligned}
\end{equation}
where ${{\bf{\hat H}}_{{\rm{eff}}}}={\bf{A}}{{\bf{\hat H}}}{{\bf{A}}^H}$ denotes the estimated channel   in the affine frequency domain. ${{\bf{\tilde H}}_{{\rm{eff}}}} = {{\bf{ H}}_{{\rm{eff}}}} - {{\bf{\hat H}}_{{\rm{eff}}}}$ is the channel estimation error matrix. Then, using the MMSE detector, the estimated data symbol, denoted by ${\bf{\hat x}}_{\rm{d}}$,  is given by ${\bf{\hat x}}_{\rm{d}} = {\bf{G\hat y}}$, 
and the MMSE equalizer $\bf{G}$ can be given by
\begin{equation}
     {\bf{G}} = {{\bf{R}}_{{\rm{x_d}}}}{\bf{\hat H}}_{{\rm{eff}}}^H{( {{{\bf{\hat H}}_{{\rm{eff}}}}{{\bf{R}}_{{\rm{x_d}}}}{\bf{\hat H}}_{{\rm{eff}}}^H + {{\bf{\tilde H}}_{{\rm{eff}}}}{{\bf{R}}_{\rm{x}}}{\bf{\tilde H}}_{{\rm{eff}}}^H +{{\bf{R}}_{\rm{n}}}} )^{ - 1}},
\end{equation}
where ${{\bf{R}}_{{\rm{x_d}}}}$ and ${{\bf{R}}_{{\rm{x}}}}$ represent the covariance matrices of ${\bf{x}}_{\rm{d}}$ and ${\bf{x}}$, respectively. ${{\bf{R}}_{\rm{n}}}$ can be expressed as ${{\bf{R}}_{\rm{n}}}={{\bf{R}}_{\rm{z}}}+\sigma _{\rm{w}}^2{{\bf{I}}_N}$.

The received signal of the $i$-th chirp subcarrier of the AFDM symbol can be represented as
\begin{equation}
\begin{aligned}
\label{deqn_ex18}
{{\bf{\hat x}}_{{\rm{d}}}}(i) = \underbrace {{{\bf{T}}{( {i,i} )}}{{\bf{x}}_{{\rm{d}}}}(i)}_{{\rm{Signal}}} + \underbrace {\sum\limits_{j \ne i} {{{\bf{T}}{( {i,j} )}}{{\bf{x}}_{\rm{d}}{(j)}}}}_{{\rm{Interference}}} + \underbrace {{{{\bf{w'}}}{( i )}}}_{{\rm{Error+Noise}}},
\end{aligned}
\end{equation}
where ${\bf{T}} = {\bf{G}}{{\bf{\hat H}}_{{\rm{eff}}}}$ is the equivalent matrix, ${\bf{w'}} = {\bf{G}}{{\bf{\tilde H}}_{{\rm{eff}}}}{\bf{x}} + {{\bf{G}}{{\bf{z}}_{\rm{mod}}}} +{\bf{Gw}}$ denotes the channel estimation error and noise matrix. The lower bound of the BER can be derived using the following theorem.

\begin{theorem}
\label{theorem2}
For MMSE detectors, if QAM is mapped to symbols using Gray code, the theoretical average BER can be approximated as a function with SINR as the independent variable. The lower bound of average BER can be obtained by using Jensen's inequality, which is given by
\begin{equation}
\begin{aligned}
\label{deqn_ex19}
{P_{\rm{AFDM}}} \ge {a_{\rm{M}}}{\rm{erfc}}( {\sqrt {\frac{{{b_{\rm{M}}}\sum\nolimits_{i = 0}^{N - 1} {\frac{{{{\bf{T}}{( {i,i} )}}}}{N}} }}{{1 - \sum\nolimits_{i = 0}^{N - 1} {\frac{{{{\bf{T}}{( {i,i} )}}}}{N}} }}} } ),
\end{aligned}
\end{equation}
where the parameter values for $a_{\rm{M}}$ and $b_{\rm{M}}$ depend on the mapping rule \cite{ref15}, and erfc-function can be expressed as ${\rm{erfc}}( x ) = \frac{2}{{\sqrt \pi  }}\int_x^\infty  {{e^{ - {t^2}}}dt} $. ${\zeta _i}$ is the SINR of the $i$-th subcarrier.
\end{theorem}

\emph{Proof: {\rm{Refer to Appendix B.}}}\hfill$\blacksquare$

\section{SIMULATION RESULTS}
In this section, we present simulation results for the proposed GCE-BEM-based channel modeling scheme for AFDM. For comparison, OTFS systems that employ GCE-BEM channel modeling are used as benchmark schemes. The number of AFDM subcarriers is $N = 256$. For OTFS, to ensure the same spectral resources are used, we set ${N_{\rm{OTFS}}} = 16$ and ${M_{\rm{OTFS}}} = 16$, which represent the number of samples in the Doppler and delay dimensions, respectively.  We consider a $3$-path channel with a maximum delay extension of ${l_{\max }} = 2$ and a maximum normalized Doppler shift of ${\alpha _{\max }} = 1$. Consider an AFDM system with a carrier frequency of ${f_{\rm{c}}} = 24 \text{GHz}$ and a subcarrier spacing of $\Delta f_{\rm{AFDM}} = 15 \text{kHz}$. The corresponding maximum speed is $675$ km/h.  In addition,  we set the BEM order $Q=4$ and the GCE-BEM oversampling coefficient $R = 2$. 
  Furthermore, Jakes Doppler spectrum is adopted, i.e, the Doppler shifts of $i$-th path is ${\alpha _i} = {\alpha _{\max }}\cos ({\theta _i})$, where ${\theta _i}$ is uniformly distributed over $[ - \pi ,\pi ]$. Denote the pilot and data signal-to-noise ratios as  ${\rm{SN}}{{\rm{R}}_{\rm{p}}} = 10\log({\left| {{x_{\rm{p}}}} \right|^2}/{\sigma ^2})$ dB and ${\rm{SN}}{{\rm{R}}_{\rm{d}}} = 10\log (\mathbb{E}\{ {| {{x_{\rm{d}}}}|^2}\} /{\sigma ^2})$ dB.

\begin{figure}[t]
\centering
\includegraphics[width=2.5in]{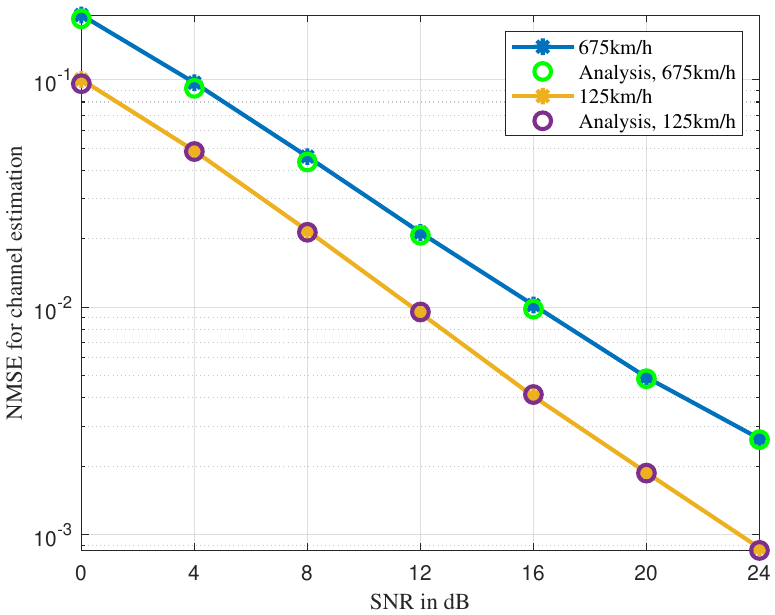}
\caption{NMSE for channel estimation at different speeds.}
\label{fig_2}
\end{figure}

Fig. \ref{fig_2} depicts the NMSE performance of the proposed GCE-BEM AFDM receiver using MMSE channel estimation with various speeds. It can be observed that the NMSE performance improves as the Doppler shift decreases. This is because that a lower Doppler shift leads to slower channel variation and reduced inter-symbol interference, thereby enhancing the accuracy of channel estimation. More importantly, the derived NMSE closely matches the Monte Carlo simulation results averaged over $10^4$ trials, validating the accuracy of our analytical derivations.

\begin{figure}[t]
\centering
\includegraphics[width=2.5in]{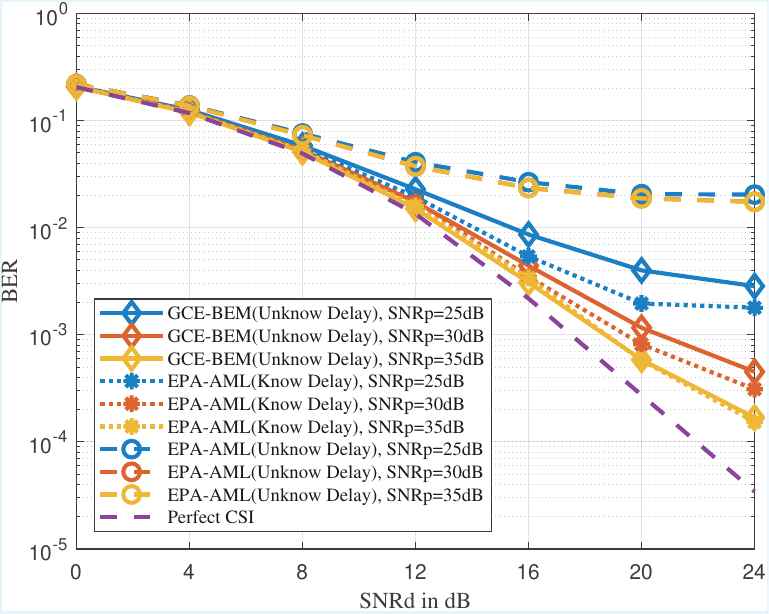}
\caption{BER performance of GCE-BEM scheme and EPA-AML scheme for the fractional Doppler shifts.}
\label{fig_3}
\end{figure}

Fig. \ref{fig_3} presents the BER performance of the proposed channel estimation schemes under both integer and fractional Doppler shifts, considering ${\rm{SNR}}_{\rm{p}} = $ 25, 30, and 35 dB. The performance is also compared with the EPA-AML scheme \cite{ref7}. 
As shown in the figure, when the channel delay is assumed to be known, the BER performance of the EPA-AML scheme is slightly better than that of the proposed GCE-BEM scheme. However, when the channel delay is unknown, the BER performance of the EPA-AML scheme degrades drastically, whereas the proposed GCE-BEM scheme effectively suppresses inter-delay interference, thereby improving performance. Since the channel delay is generally unknown in practice, the proposed GCE-BEM scheme outperforms  the EPA-AML scheme.


\begin{figure}[t]
\centering
\includegraphics[width=2.5in]{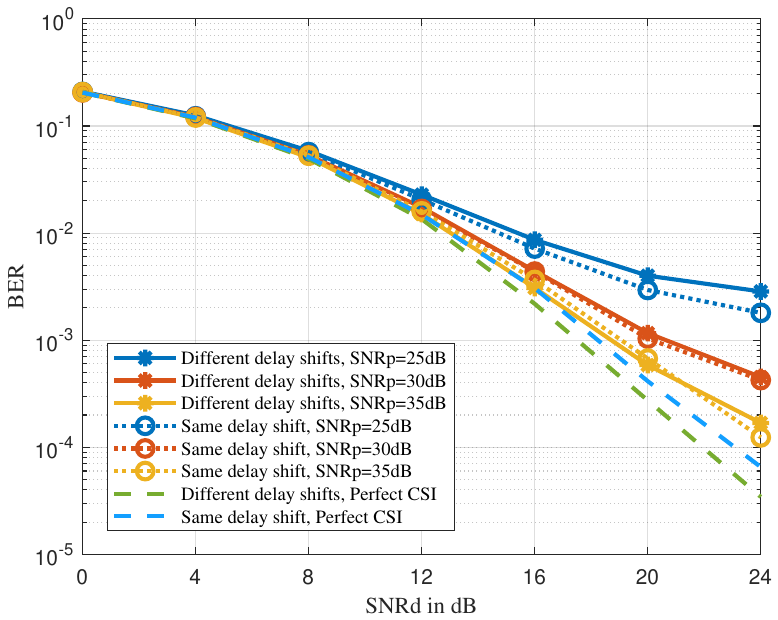}
\caption{BER performance of GCE-BEM scheme for the different and same delay shifts, fractional Doppler.}
\label{fig_5}
\end{figure}

Based on the estimation error, we further present the BER performance of the GCE-BEM scheme under both different and same delay shifts, as shown in Fig. \ref{fig_5}.  It can be seen that the proposed GCE-BEM method estimates the time-domain channel by approximating the time-varying channel as a linear combination of mutually orthogonal basis functions, thus effectively avoiding the inter-Doppler interference (IDoI) in the affine frequency domain channel. Notably, even when paths have same delay shifts, the GCE-BEM-based AFDM system maintains robust BER performance.

\begin{figure}[t]
\centering
\includegraphics[width=2.5in]{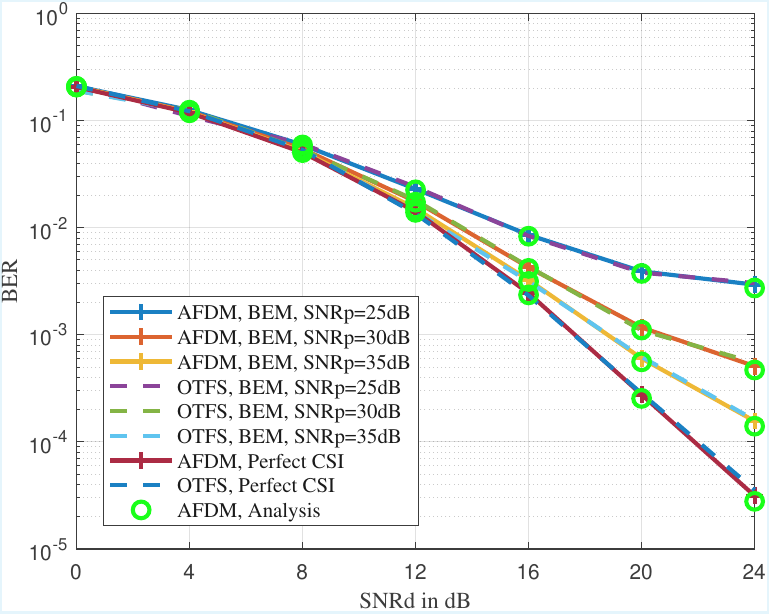}
\caption{BER performance of AFDM and OTFS systems based on GCE-BEM, fractional Doppler.}
\label{fig_6}
\end{figure}

Fig. \ref{fig_6} compares the BER performance of OTFS and AFDM systems under GCE-BEM channel modeling for different values of ${\rm{SNR}}_{\rm{p}}$. The embedded pilot structure proposed in \cite{ref14} is considered OTFS system. 
As observed in the figure, increasing ${\rm{SNR}}_{\rm{p}}$ from 25 to 35 dB leads to significant improvements in BER performance due to reduced channel estimation errors.  It can be observed   that the BER performance of the AFDM and OTFS systems is comparable. However, it is worth noting that AFDM generally requires lower channel estimation overhead \cite{ref7}.
More importantly, the analytical BER closely matches the results of Monte Carlo simulations, validating the accuracy of the proposed analytical model.

\section{CONCLUSION}
In this paper, an AFDM receiver scheme based on GCE-BEM was proposed to reduce the computational complexity of channel estimation and solve the severe IDoI induced by fractional Doppler dispersion. We derived the GCE-BEM-based channel representation in the DAFT domain. Then, AFDM system model based on a two-pilot embeded structure was introduced. In particular, we further analyzed the NMSE and BER based on our estimated channel channel estimation. The simulation results validated the accuracy of the theoretical analysis and demonstrated the superiority of the AFDM system based on the GCE-BEM method over the benchmark schemes, underscoring its great potential for practical applications.

{\appendix \subsection{Proof of Theorem \ref{theorem1}.}
\label{appendix_1}
 ${{\bf{R}}_{\rm{d}}} $ is the data covariance matrix with its expression given by
\begin{equation}
\begin{aligned}
\label{deqn_ex20}
{{\bf{R}}_{\rm{d}}}& = \mathbb{E}\{ {{{\bf{\Psi}} _{\rm{d}}}{\bf{g}}{{\bf{g}}^H}{\bf{\Psi}} _{\rm{d}}^H} \}\\
& = {\bf{\Upsilon }}( {{{\bf{J}}_{Q + 1}} \otimes ( {( {{\bf{F}}{{\bf{A}}^{\rm{H}}}} ){{\bf{R}}_{{\rm{x_d}}}}{{( {{\bf{F}}{{\bf{A}}^{\rm{H}}}} )}^H}} )} ) \odot ( { {\bf{\Xi}} {{\bf{R}}_{\rm{g}}}{{ {\bf{\Xi}}^H }}} ){{\bf{\Upsilon }}^H},
\end{aligned}
\end{equation}
where ${\bf{\Upsilon }} = [{{\bf{\Upsilon }}_0},{{\bf{\Upsilon }}_1}, \ldots ,{{\bf{\Upsilon }}_Q}] \in {{\mathbb{C}}^{N \times N(Q + 1)}}$ is a deterministic matrix associated with the basis function, and ${{\bf{\Upsilon }}_{q}}={\bf{A}}{\rm{diag}}\{ {{\bf{b}}_q}\}{{\bf{F}}^{\rm{H}}}$ denotes basis function correlation matrix. ${{\bf{J}}_{Q + 1}}\in {{\mathbb{C}}^{(Q + 1) \times (Q + 1)}}$ denotes an all-ones matrix. ${\bf{\Xi}}$ can be expressed as ${\bf{\Xi}} = {{{\bf{I}}_{Q + 1}} \otimes {{\bf{F}}_L}}$, and $\odot$ denotes the Hadamard product.

By defining ${{\bf{e}}_{{\rm{mod}} ,l}} = {\left[ {{e_{\rm{mod} }}(0,l),{e_{\rm{mod} }}(1,l), \ldots ,{e_{\rm{mod} }}(N - 1,l)} \right]^T}$ as the channel model error of $l$-th path, and ${e_{\rm{mod} }}(n,l)$ denotes the model error. ${{\bf{z}}_{\rm{mod} }}$ can be further expressed as ${{\bf{z}}_{{\rm{mod}} }} = \sum\nolimits_{l = 0}^L {{\bf{A}}{\rm{diag}}\{ {{\rm{circshift}}( {{\bf{s}},l} )} \}{{\bf{e}}_{{\rm{mod}} ,l}}}$.

We define ${\bf{s}} = {{\bf{s}}_{\rm{p}}} + {{\bf{s}}_{\rm{d}}}$, where ${{\bf{s}}_{\rm{p}}}$ and ${{\bf{s}}_{\rm{d}}}$ are time-domain signals corresponding to pilot and data, respectively. Since ${\bf{g}}$, ${\bf{x}}$, and ${\bf{w}}$ are independent of each other with zero mean, the autocorrelation matrix of ${{\bf{z}}_{\rm{mod} }}$ can be expressed as
\begin{equation}
\begin{aligned}
\label{deqn_ex22}
{{\bf{R}}_{\rm{z}}}& = \mathbb{E}\{ {{{\bf{z}}_{\rm{mod} }}{\bf{z}}_{\rm{mod} }^H} \}\\
& = \mathbb{E}\{ {{{\bf{z}}_{{\rm{mod}} ,{\rm{p}}}}{\bf{z}}_{{\rm{mod}} ,{\rm{p}}}^H} \} + \mathbb{E}\{ {{{\bf{z}}_{{\rm{mod}} ,{\rm{d}}}}{\bf{z}}_{{\rm{mod}} ,{\rm{d}}}^H} \}\\
& = \sum\nolimits_{l = 0}^L {{\bf{A}}{{\bf{c}}_{l,{\rm{p}}}}{{\bf{R}}_{{\rm{mod}} ,l}}{\bf{c}}_{l,{\rm{p}}}^H{{\bf{A}}^H}} \\
&+{{\bf{A}}( {{{\bf{c}}_{{\rm{A}},l}} {{\bf{R}}_{{\rm{x_d}}}} {{\bf{c}}_{{\rm{A}},l}^H} \odot {{\bf{R}}_{{\rm{mod}} ,l}}} ){{\bf{A}}^H}},
\end{aligned}
\end{equation}
where ${{\bf{c}}_{l,{\rm{p}}}} = {\rm{diag\{ circshift(}}{{\bf{s}}_{\rm{p}}}{\rm{,}}l{\rm{)\} }}$ are diagonal matrices associated with pilot, and ${{\bf{c}}_{{\rm{A}},l}} = {\rm{circshift}}( {{{\bf{A}}^H},l} )$ denotes the cyclic shift matrix of ${\bf{A}}$. Defining ${\bf{\Phi}}  = {{\bf{I}}_N} - {\bf{B}}{({{\bf{B}}^H}{\bf{B}})^{ - 1}}{{\bf{B}}^H}$, ${{\bf{R}}_{{\rm{mod} },l}} = \mathbb{E}\{ {{{\bf{e}}_{{\rm{mod}},l }}{\bf{e}}_{{\rm{mod}},l }^H} \} = {\bf{\Phi}} {{\bf{R}}_{{\rm{hh}},l}}{{\bf{\Phi}} ^H}$\cite{ref14} is the covariance matrix of the $l$-th path model error ${{\bf{e}}_{{\rm{mod}} ,l}}$.

Then, by substituting \eqref{deqn_ex12} into formula ${\bf{\tilde g}} = {\bf{g}} - {\bf{\hat g}}$, the covariance matrix of BEM estimation error can be expressed as
\begin{equation}
\begin{aligned}
\label{deqn_ex25}
{{\bf{R}}_{\tilde{{\rm{g}}}}} = \mathbb{E}\{ {{\bf{\tilde g}}{\bf{\tilde g}}^H} \} = {{\bf{R}}_{\rm{g}}} - {{\bf{V}}_{{\rm{MMSE}}}}{{\bf{\Psi}} _{\rm{p,p}}}{\bf{R}}_{\rm{g}}^H,
\end{aligned}
\end{equation}
where ${{\bf{V}}_{\rm{MMSE}}} = {{\bf{R}}_{\rm{g}}}{\bf{\Psi}} _{{\rm{p}},{\rm{p}}}^H({{\bf{\Psi}} _{{\rm{p}},{\rm{p}}}}{{\bf{R}}_{\rm{g}}}{\bf{\Psi}} _{{\rm{p}},{\rm{p}}}^H + {{\bf{R}}_{{\rm{d}},{\rm{p}}}} + {{\bf{R}}_{{\rm{z}},{\rm{p}}}} + {{\bf{R}}_{{\rm{w}},{\rm{p}}}})^{ - 1}$ is linear estimation matrix.

Finally, the total NMSE of system channel estimation is expressed as
\begin{equation}
\begin{aligned}
\label{deqn_ex26}
{\rm{NMSE}}_{{\rm{total}}}& = \frac{{{\rm{trace(}}\mathbb{E}\{ {{\bf{e}}_{\rm{mod} }}{\bf{e}}_{\rm{mod} }^H + ({\bf{\bar h}} - {\bf{\hat h}}){{({\bf{\bar h}} - {\bf{\hat h}})}^H}\} )}}{{{\rm{trace(}}\mathbb{E}\{ {\bf{h}}{{\bf{h}}^H}\} )}}\\
&= \frac{{\sum\nolimits_{l = 0}^L {{\rm{trace\{ }}{\bf{\Phi}} {{\bf{R}}_{{\rm{hh}},l}}{\rm{\} }}} } +{{{\rm{trace\{ }}{\bf{\Theta}}{{\bf{R}}_{\tilde{{\rm{g}}}}}  { {\bf{\Theta}}^H }{\rm{\} }}}}}{{{\rm{trace\{ }}{{\bf{R}}_{{\rm{hh}}}}{\rm{\} }}}} ,
\end{aligned}
\end{equation}
where ${{\bf{e}}_{\rm{mod}}} = {\bf{h}} - {\bf{\bar h}}$ is the modeling error vector, and $\mathbb{E}\{ {{\bf{e}}_{\rm{mod} }}{\bf{e}}_{\rm{mod} }^H\} = {\sum\nolimits_{l = 0}^L {{\rm{trace\{ }}{\bf{\Phi}} {{\bf{R}}_{{\rm{hh}},l}}{\rm{\} }}} }$\cite{ref14}.  ${\rm{trace}}\{  \cdot \} $ denotes the trace of a matrix, and ${{\bf{R}}_{{\rm{hh}},l}}$ is the autocorrelation matrix of channel gain of the $l$-th path. ${\bf{\bar h}} = {\bf{\Theta}}{\bf{g}}$ is the BEM modeling channel gain coefficient vector, and ${\bf{\hat h}} = {\bf{\Theta}}{\bf{\hat g}}$ is the estimated channel.

\subsection{Proof of Theorem \ref{theorem2}.}
According to \eqref{deqn_ex18}, the total power of the received signal can be expressed as
\begin{equation}
\begin{aligned}
\label{deqn_ex27}
&\mathbb{E}\{ {{{|{{{\bf{\hat x}}_{{\rm{d}}}}(i)}|}^2}} \}\\
& = {{\bf{G}}_i}( {{{\bf{\hat H}}_{{\rm{eff}}}}{{\bf{R}}_{{\rm{x_d}}}}{\bf{\hat H}}_{{\rm{eff}}}^H + {{\bf{\tilde H}}_{{\rm{eff}}}}{{\bf{R}}_{{\rm{x}}}}{\bf{\tilde H}}_{{\rm{eff}}}^H + {{\bf{R}}_{\rm{z}}}+\sigma _{\rm{w}}^2{{\bf{I}}_N}} ){\bf{G}}_i^H\\
& = {\varepsilon _{{\rm{x_d}}}}{{\bf{G}}_i}{{\bf{\hat H}}_{{\rm{eff}}}}(i),
\end{aligned}
\end{equation}
where ${\varepsilon _{{\rm{x_d}}}}$ is the power of the transmitter data signal. Therefore, it can be further obtained from the above formula
\begin{equation}
\begin{aligned}
\label{deqn_ex28}
{{\bf{T}}{( {i,i} )}} = {{\bf{G}}_i}{{\bf{\hat H}}_{{\rm{eff}}}}(i) = \mathbb{E}\{ {{{| {{{\bf{\hat x}}_{{\rm{d}}}}(i)} |}^2}} \}/{\varepsilon _{{\rm{x_d}}}}.
\end{aligned}
\end{equation}

Then, the SINR of the $i$-th subcarrier can be expressed as
\begin{equation}
\begin{aligned}
\label{deqn_ex29}
{\zeta _i} = \frac{{{\varepsilon _{{\rm{x_d}}}}{\bf{T}}^2{( {i,i} )}}}{{\mathbb{E}\{ {{{| {{{\bf{\hat x}}_{{\rm{d}}}}(i)} |}^2}} \} - {\varepsilon _{{\rm{x_d}}}}{\bf{T}}^2{( {i,i} )}}} = \frac{{{{\bf{T}}{( {i,i} )}}}}{{1 - {{\bf{T}}{( {i,i} )}}}}.
\end{aligned}
\end{equation}

By taking the average BER of each subcarrier, the BER expression of the AFDM system can be obtained as
\begin{equation}
\begin{aligned}
\label{deqn_ex30}
{P_{{\rm{AFDM}}}} = \frac{1}{N}\sum\limits_{i = 0}^{N - 1} {{a_{\rm{M}}}{\rm{erfc}}( {\sqrt {\frac{{{b_{\rm{M}}}{{\bf{T}}{( {i,i} )}}}}{{1 - {{\bf{T}}{( {i,i} )}}}}} } )}.
\end{aligned}
\end{equation}
In \cite{ref16}, it has shown that the value range of the independent variable $\bf{x}$ when function $\varphi ( x ) = {\rm{erfc}}( {\sqrt {\frac{{{b_{\rm{M}}}x}}{{1 - x}}} } )$ is a convex function under each mapping rule, and the convexity holds in the region of high SINR. Therefore, by using the convexity and Jensen's inequality, the lower bound of \eqref{deqn_ex30} can be proven as \eqref{deqn_ex19}.

 
%
}

\end{document}